\documentclass[a4paper,11pt]{article}
\usepackage{pos}
\usepackage{bm}
\usepackage{subcaption}

\title{A novel framework for spectral density reconstruction via quadrature-based Laplace inversion}
\ShortTitle{Spectral density reconstruction via quadrature-based Laplace inversion}

\author[a]{Marco Aliberti}
\author[a]{Francesco Di Renzo}
\author[a]{Petros Dimopoulos}
\author*[a]{Demetrianos Gavriel}

\affiliation[a]{Dipartimento di Scienze Matematiche, Fisiche e Informatiche, Universita di Parma and INFN, Gruppo Collegato di Parma, I-43100 Parma, Italy}

\emailAdd{marco.aliberti@unipr.it}
\emailAdd{francesco.direnzo@unipr.it}
\emailAdd{petros.dimopoulos@unipr.it}
\emailAdd{demetrianos.gavriel@unipr.it}

\abstract{In this work, we explore a numerical approach for performing the inverse Laplace transformation, with an emphasis on achieving stability and robustness under noisy conditions. Our quadrature-based method integrates reparameterization, data smoothing, and optimization techniques to regularizing ill-conditioned systems. Together, these elements enable consistency checks that enhance the reliability of the inversion process. Through a series of controlled tests on toy models, we demonstrate the stability and effectiveness of the method in the presence of noise. Using mock data, we approximate spectral densities from Euclidean correlators, generating smoothed and stable results that accurately reproduce the correlator behavior, particularly at large Euclidean times. We conclude by discussing prospects for applications to actual lattice QCD data.}

\FullConference{The 42nd International Symposium on Lattice Field Theory (LATTICE2025)\\
2-8 November 2025\\
Tata Institute of Fundamental Research, Mumbai, India\\}

\begin{document}

\maketitle

\section{Introduction}

Inverse problems are ubiquitous in physics, with a notable example being the reconstruction of spectral densities from Euclidean correlators in lattice QCD via an inverse Laplace transform. This problem is severely ill-posed due to discretization, statistical noise, and finite-volume effects. Standard approaches, such as Bayesian, maximum entropy, or Backus--Gilbert methods, stabilize the inversion by introducing priors or resolution functions, at the cost of some degree of model dependence.

In these proceedings, we explore an alternative strategy that reformulates the inverse Laplace transform as a linear system using Gauss-type quadrature and a systematic reparameterization. By comparing reconstructions across reparameterization scales, stable inversion regions can be identified in a data-driven manner. We outline the method, validate it on analytic toy models, and present first results for mock lattice correlators.

\section{Inverse Laplace Problem}

The Laplace transform of a function $f(t)$ that vanishes sufficiently fast at infinity is defined as
\begin{equation}
F(s) = \int_{0}^{\infty} dt \, e^{-s t} f(t),
\label{eq:Laplace_transformation}
\end{equation}
with $\Re(s)>0$. The inverse Laplace transform is formally given by the Bromwich integral,
\begin{equation}
f(t) = \frac{1}{2\pi i} \int_{\gamma - i\infty}^{\gamma + i\infty} ds \, e^{s t} F(s),
\end{equation}
where $\gamma$ lies to the right of all singularities of $F(s)$.

In practice, analytic inversion is rarely feasible. Classical numerical inversion techniques (see, for example, \cite{DaviesMartin,HansenInverse}) typically assume knowledge of $F(s)$ as a continuous function, often in the complex plane. In many physical applications, including lattice QCD, the Laplace-space data are instead available only at a finite set of real points and are affected by statistical noise, making the inverse problem highly ill-posed.

The central observation of this work is that the Laplace transform itself is an integral transform that can be discretized systematically. Approximating the integral using Gaussian quadrature allows the inverse Laplace problem to be recast as a linear algebra problem relating discrete samples of $F(s)$ to discrete samples of $f(t)$. This formulation relies only on real-axis data and avoids assumptions about analytic continuation. No explicit priors on the spectral density are introduced beyond those inherent to the Laplace representation.

\section{Quadrature-Based Formulation}

\subsection{Gauss--Laguerre Quadrature}

For integrals over the semi-infinite interval $[0,\infty)$ with an exponential weight, Gauss--Laguerre quadrature provides an efficient approximation,
\begin{equation}
\int_{0}^{\infty} dx \, e^{-x} g(x) \approx \sum_{j=1}^{n} w_j g(x_j),
\label{eq:Gauss-Laguerre_quadrature}
\end{equation}
where $x_j$ are the roots of the Laguerre polynomial $L_n(x)$ and $w_j$ are the corresponding quadrature weights \cite{DavisRabinowitz, hildebrand1987}.

Applying this quadrature to Eq.~(\ref{eq:Laplace_transformation}) allows for a numerical representation of the Laplace transform when $F(s)$ is known only at a finite set of points. The quadrature nodes depend solely on the chosen order $n$ and are independent of the functional form of the inverse function $f(t)$, making the approach well suited for problems with limited prior information. Gauss--Laguerre quadrature is particularly appropriate here, as its exponential weight matches the structure of the Laplace kernel and minimizes quadrature error for smooth integrands.

\subsection{Reparameterization}

We introduce a reparameterization scale $t_0$ to the Laplace transformation by defining $t = t_0 t'$. This yields 
\begin{equation}
F(s) = t_0 \int_{0}^{\infty} dt' \, e^{-s t_0 t'} f(t_0 t').
\end{equation}
Applying Gauss--Laguerre quadrature to this expression for a discrete set of Laplace points $s_i$ gives
\begin{equation}
F(s_i) \approx t_0 \sum_{j=1}^{n} w_j\, e^{-t_j(t_0 s_i-1)} f(t_0 t_j).
\end{equation}
This expression can be written as a linear system,
\begin{equation}
\bm{b} = \bm{A} \cdot \bm{x},
\end{equation}
where $\bm{b}$ contains the known values $F(s_i)$, matrix $\bm{A}$ encodes the exponential kernel and quadrature weights, and $\bm{x}$ represents the unknown values of $f(t_0 t_j)$.

For fixed $t_0$, the inversion reduces to solving a generally ill-conditioned linear system. The scale $t_0$ controls both the effective resolution and the numerical conditioning of the problem. Rather than fixing this scale in advance, we exploit its variation to identify stable inversion regimes.

\subsection{Stability Criterion}

In the absence of an analytic solution, the optimal reparameterization scale cannot be determined by direct comparison. We therefore define a stability indicator based on the relative variation between solutions obtained at consecutive scales,
\begin{equation}
R_k = \frac{\|\bm{x}_{k+1}-\bm{x}_{k}\|}{\|\bm{x}_{k}\|}.
\label{eq:consecutive_solutions_reparameterization}
\end{equation}
A minimum in $R_k$ identifies a region where the reconstructed solution is least sensitive to variations in $t_0$, indicating a stable inversion window.

\begin{figure}[htbp]
    \centering
    \includegraphics[width=0.5\textwidth]{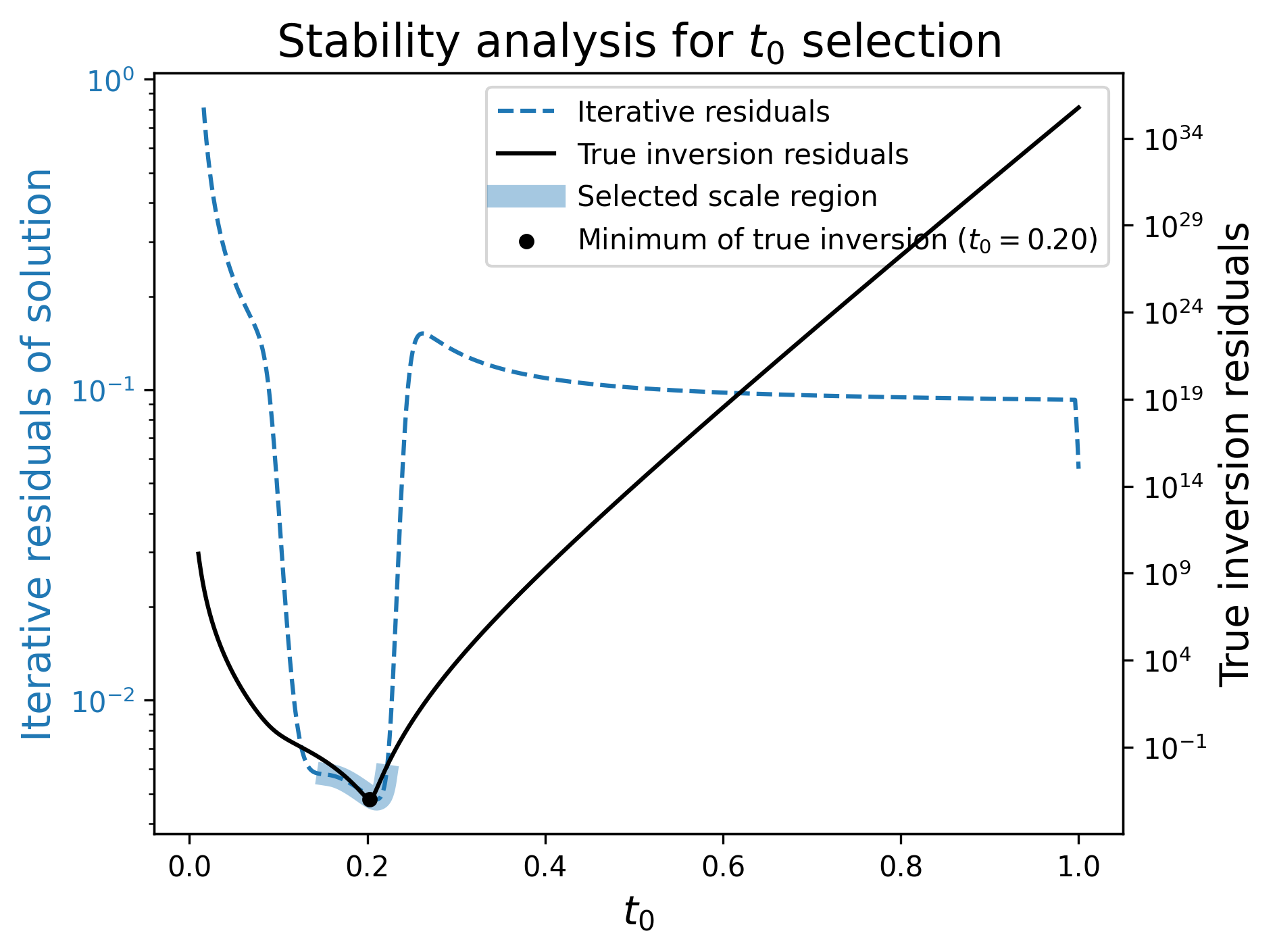}
    \caption{
    Stability analysis for choosing the reparameterization scale $t_0$ in the $F(s)=1/s^2$ analytic test case. The blue dashed curve (left axis) shows relative differences between consecutive $t_0$ reconstructions (Eq.~\ref{eq:consecutive_solutions_reparameterization}), and the black curve (right axis) shows the true inversion residual. Both have a clear minimum near $t_0=0.2$. The shaded band marks the stable region, and the black marker shows the residual minimum.
    }
    \label{fig:scale_selection}
\end{figure}

In controlled tests where the exact inverse is known, this criterion agrees well with the minimum of the true reconstruction error, as illustrated in Fig.~\ref{fig:scale_selection} for the case $F(s)=1/s^2$ (studied further in the next section). The use of consecutive scales provides a local probe of stability, avoiding sensitivity to long-range fluctuations in scale space. This gives a practical, data-driven method for scale selection that does not rely on external priors.

\section{Tests on Analytic Toy Models}

To validate our method, we consider the analytic test case
\begin{equation}
F(s) = \frac{1}{s^2},
\end{equation}
whose exact inverse is $f(t)=t$.

\begin{figure}[htbp]
    \centering

    \begin{subfigure}[t]{0.48\textwidth}
        \centering
        \includegraphics[width=\linewidth]{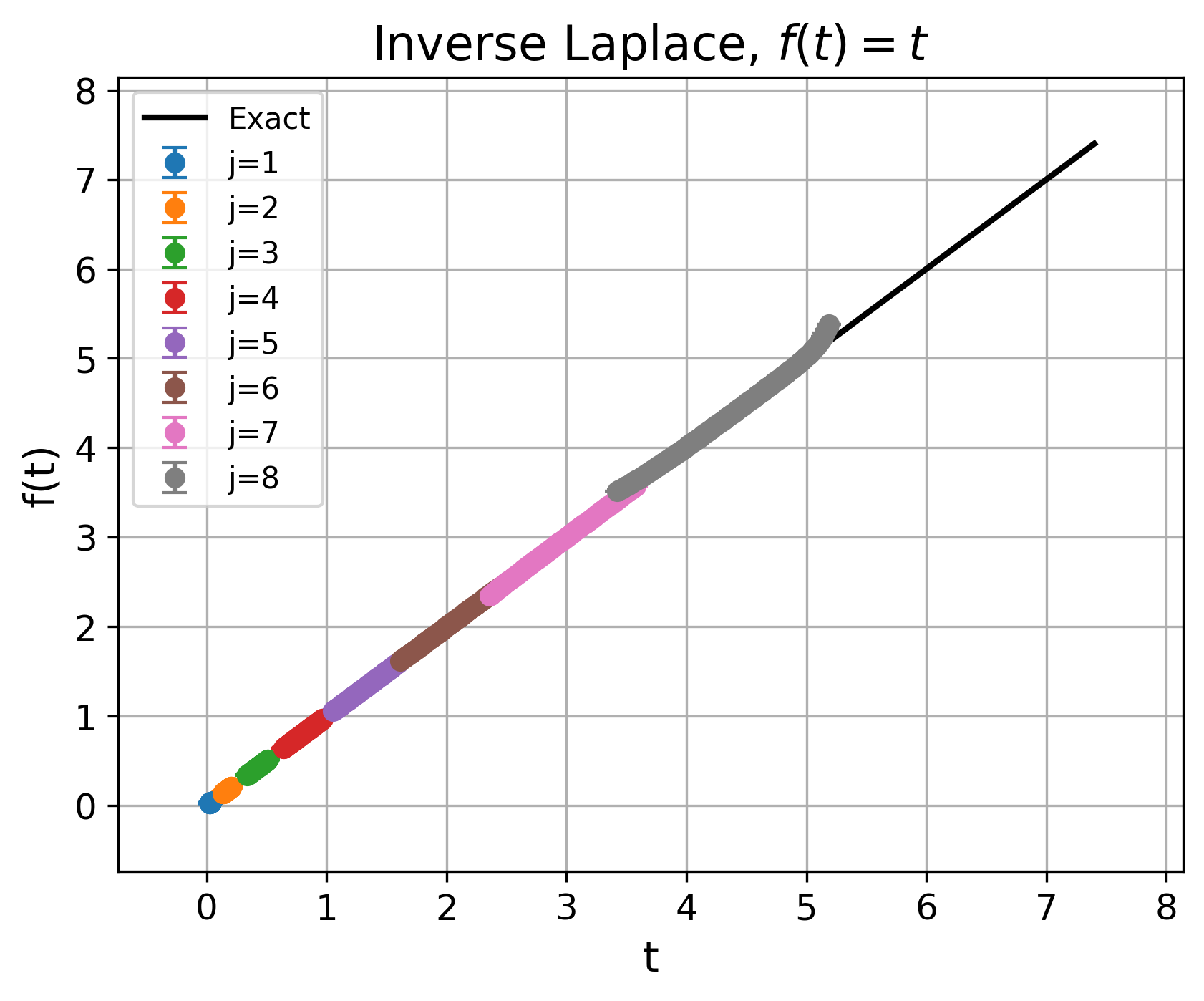}
        \caption{
        Inverse Laplace reconstruction of the analytic function $f(t)=t$ from eight Laplace-space samples of $F(s)=1/s^{2}$ in the interval $s \in [3.5,6.5]$, in the absence of noise. Colored markers correspond to reconstructed values at the Gauss–Laguerre quadrature nodes $t_0 t_j$ for different indices $j$, while the black line shows the exact solution.}
        \label{fig:inverse_laplace}
    \end{subfigure}
    \hfill
    \begin{subfigure}[t]{0.48\textwidth}
        \centering
        \includegraphics[width=\linewidth]{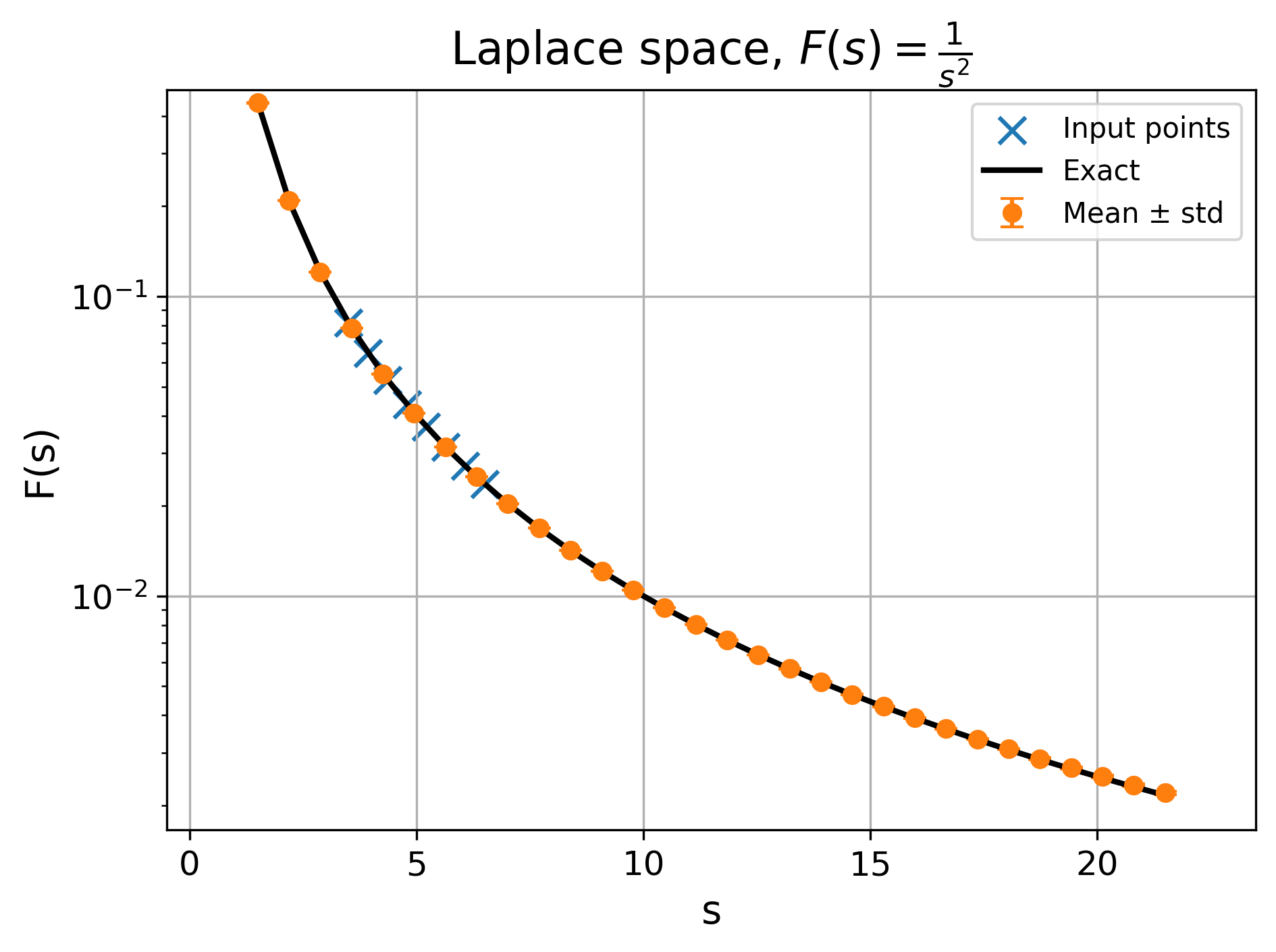}
        \caption{
        Reconstruction of the Laplace-space function $F(s)$ from the inverted values shown in panel \ref{fig:inverse_laplace}. Blue crosses indicate the original input data points, while orange circles show reconstructed and extrapolated values obtained from the inverse solution. The black curve represents the exact result $F(s)=1/s^2$.
        }
        \label{fig:reconstruction_extrapolation}
    \end{subfigure}
    
    \caption{Numerical inversion and reconstruction test for $F(s)=1/s^{2}$ with analytic inverse $f(t)=t$, performed in the absence of noise.}
    \label{fig:laplace_test}
\end{figure}

We sample $F(s)$ at eight evenly spaced points in $s\in[3.5,6.5]$ and perform the quadrature-based inversion using $n=8$ Gauss--Laguerre nodes for 100 values of the reparameterization scale in the range $t_0\in[0.15,0.23]$, selected via the stability criterion introduced earlier. The reconstructed values $f(t_0 t_j)$, shown in Fig.~\ref{fig:inverse_laplace}, exhibit a clear stability window in which the numerical solution accurately reproduces the analytic result.

From the reconstructed inverse, we recompute $F(s)$ at 30 points in the extended interval $s\in[1.5,21.5]$. As shown in Fig.~\ref{fig:reconstruction_extrapolation}, the stable inversion enables accurate reconstruction and extrapolation of the Laplace-space function.

\begin{figure}[htbp]
    \centering

    %---------------- Column 1 ----------------
    \begin{subfigure}[t]{0.32\textwidth}
        \centering
        \includegraphics[width=\linewidth]{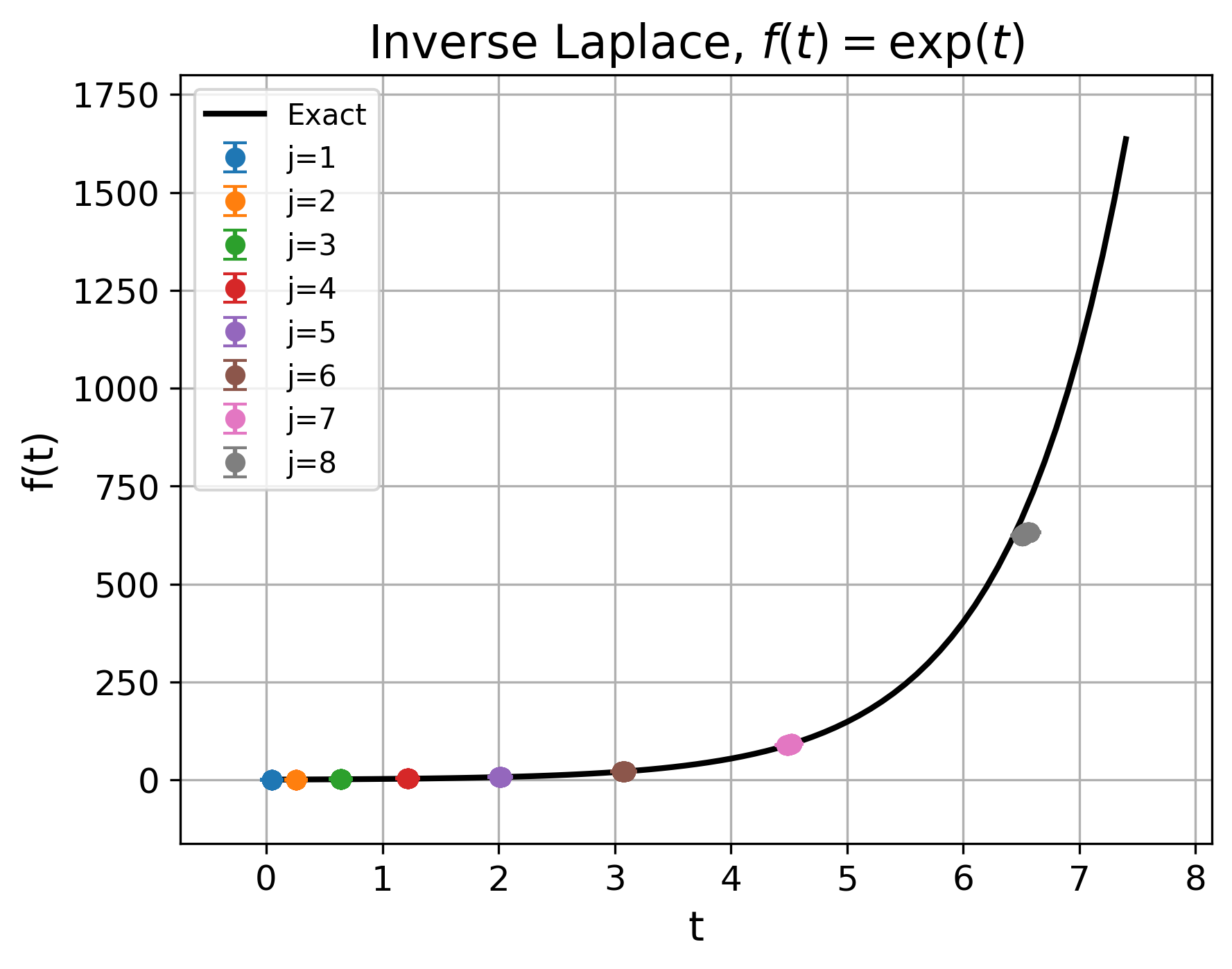}
        \vspace{0.5em}
        \includegraphics[width=\linewidth]{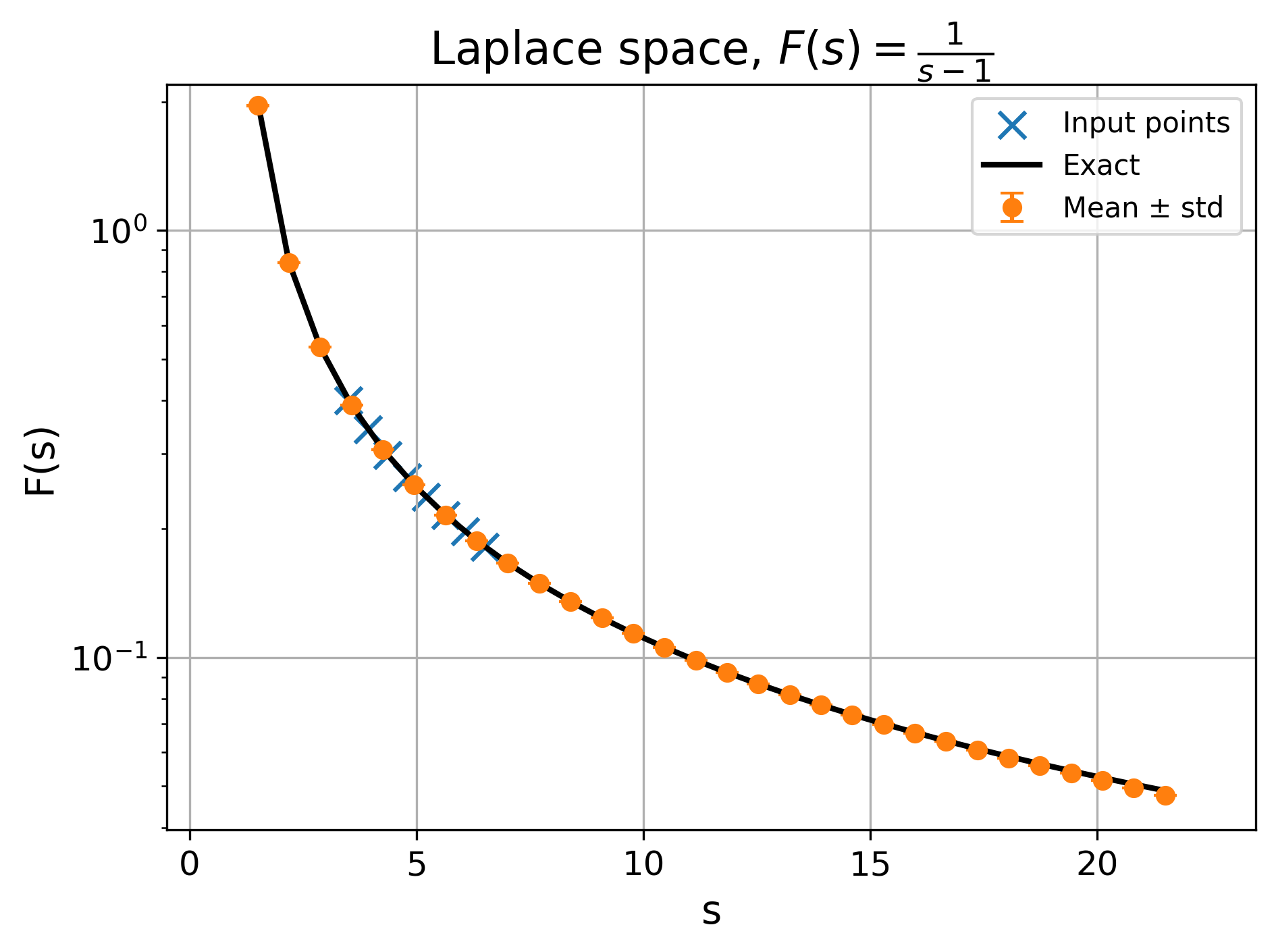}
    \end{subfigure}
    \hfill
    %---------------- Column 2 ----------------
    \begin{subfigure}[t]{0.32\textwidth}
        \centering
        \includegraphics[width=\linewidth]{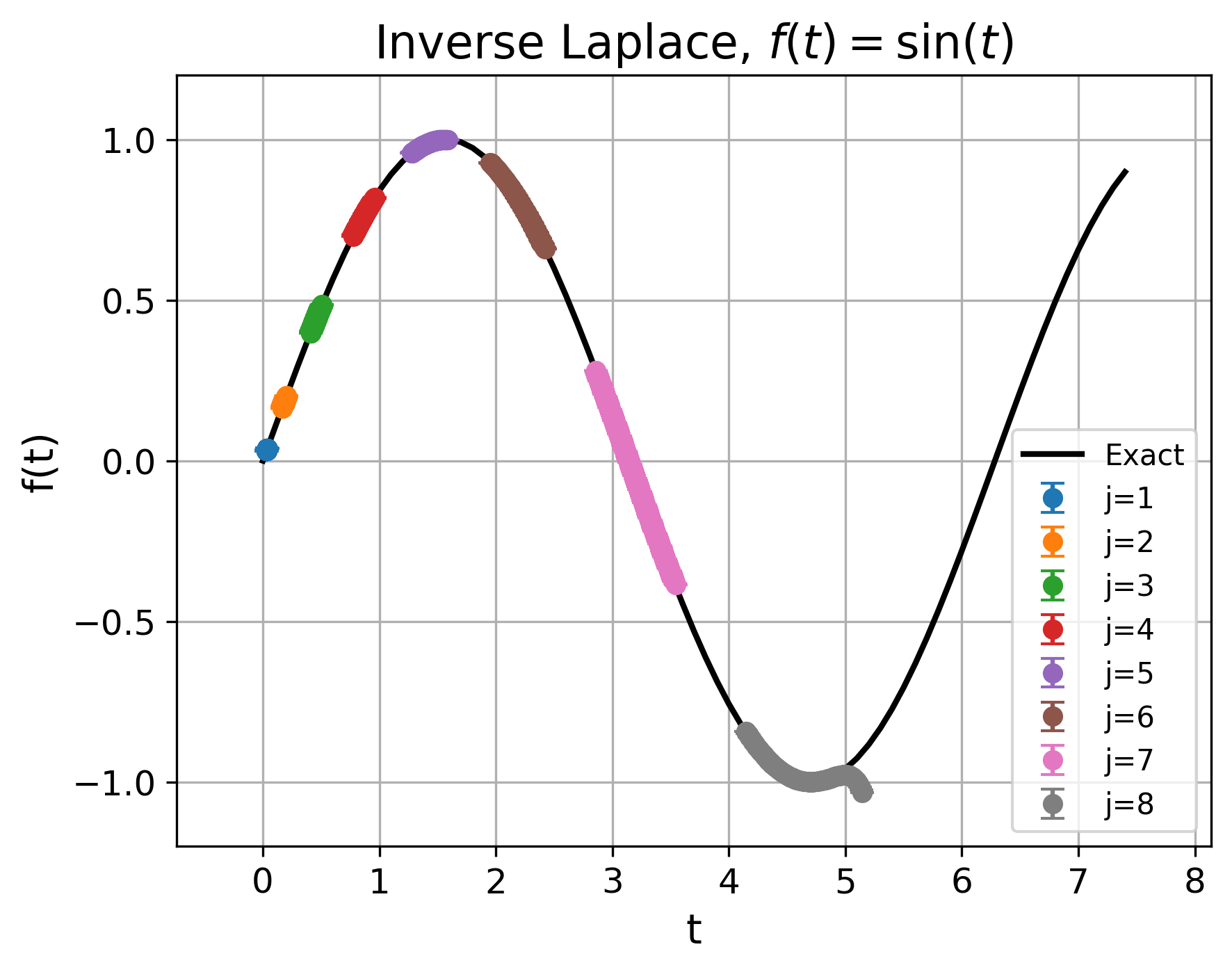}
        \vspace{0.5em}
        \includegraphics[width=\linewidth]{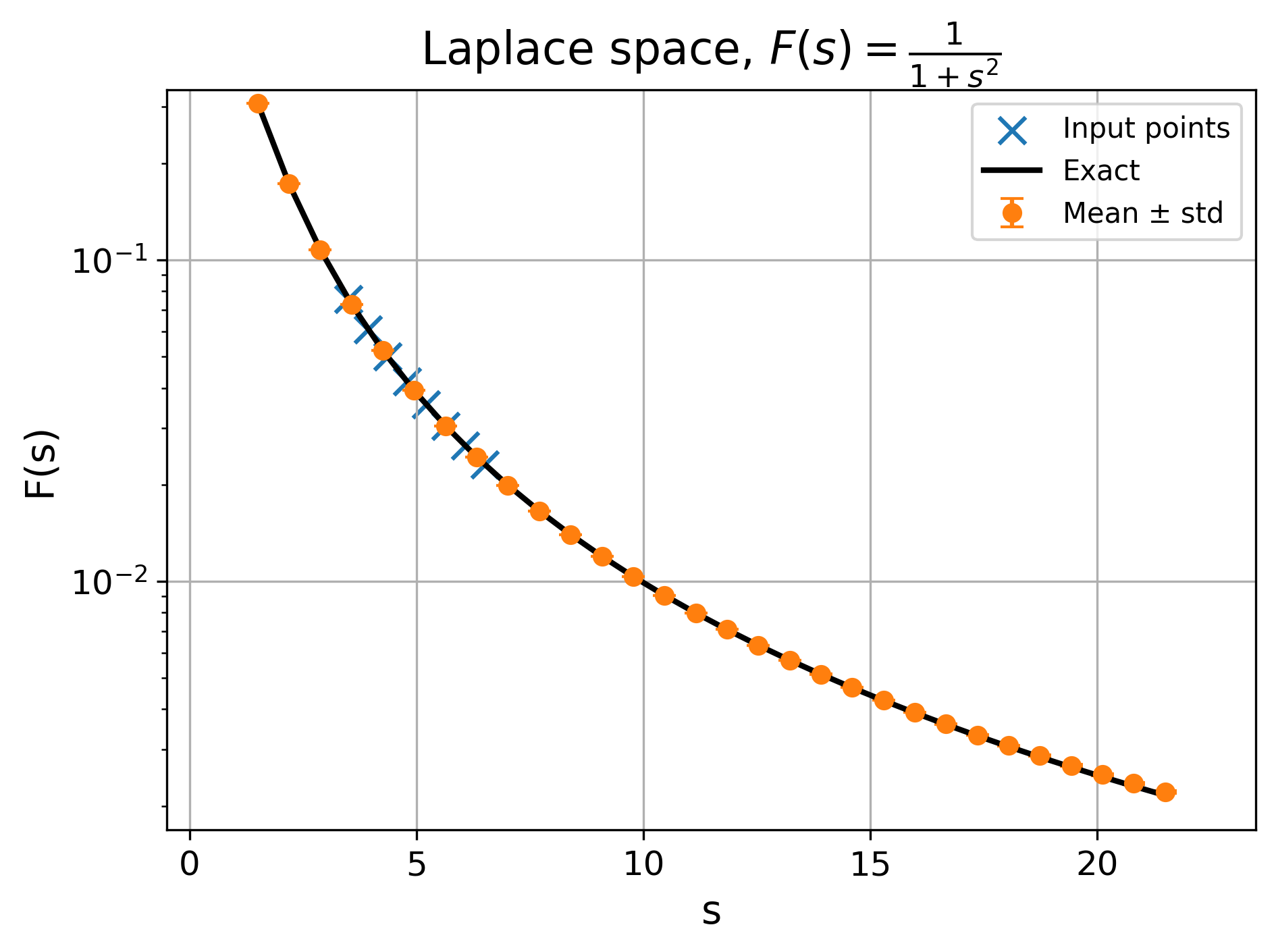}
    \end{subfigure}
    \hfill
    %---------------- Column 3 ----------------
    \begin{subfigure}[t]{0.32\textwidth}
        \centering
        \includegraphics[width=\linewidth]{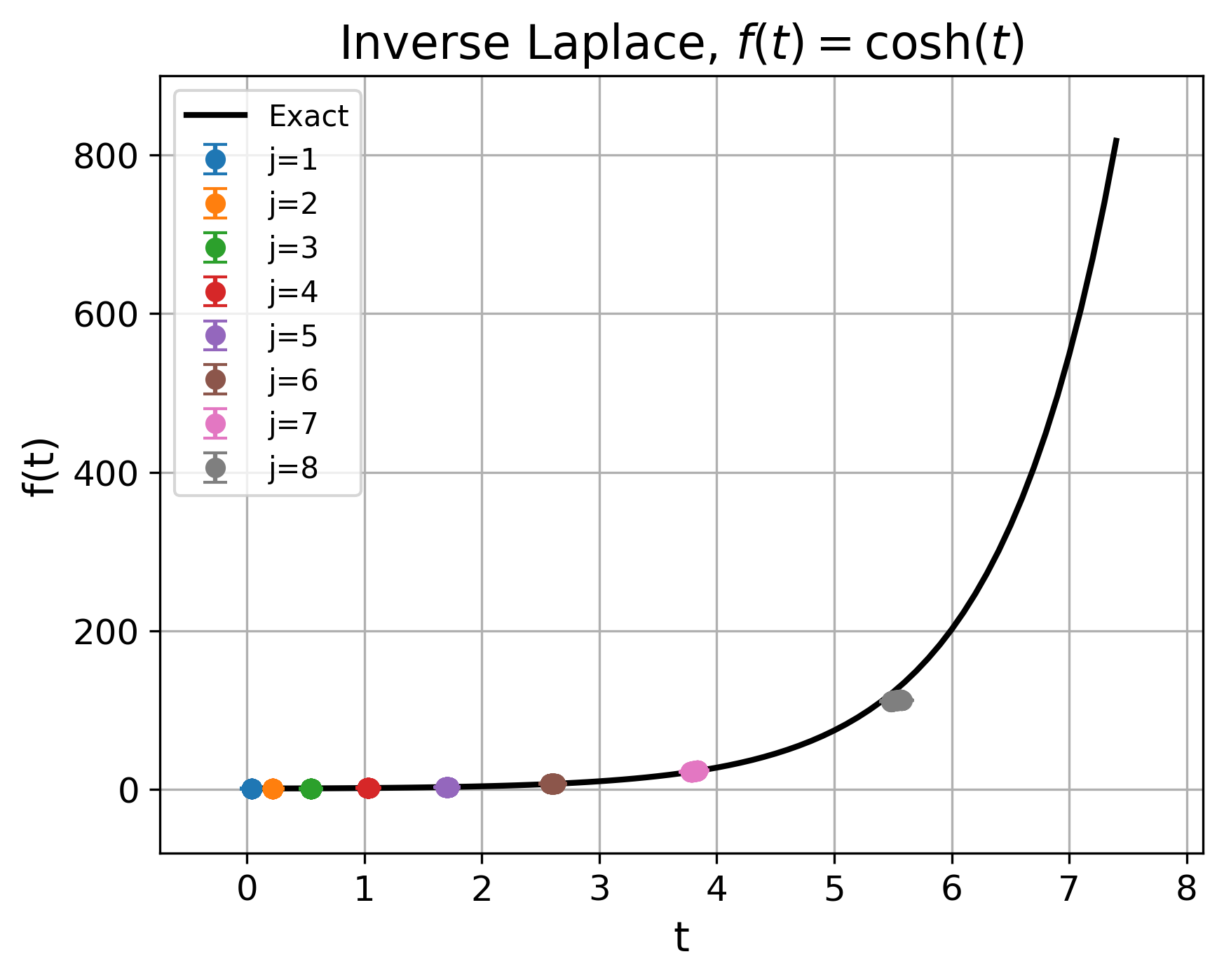}
        \vspace{0.5em}
        \includegraphics[width=\linewidth]{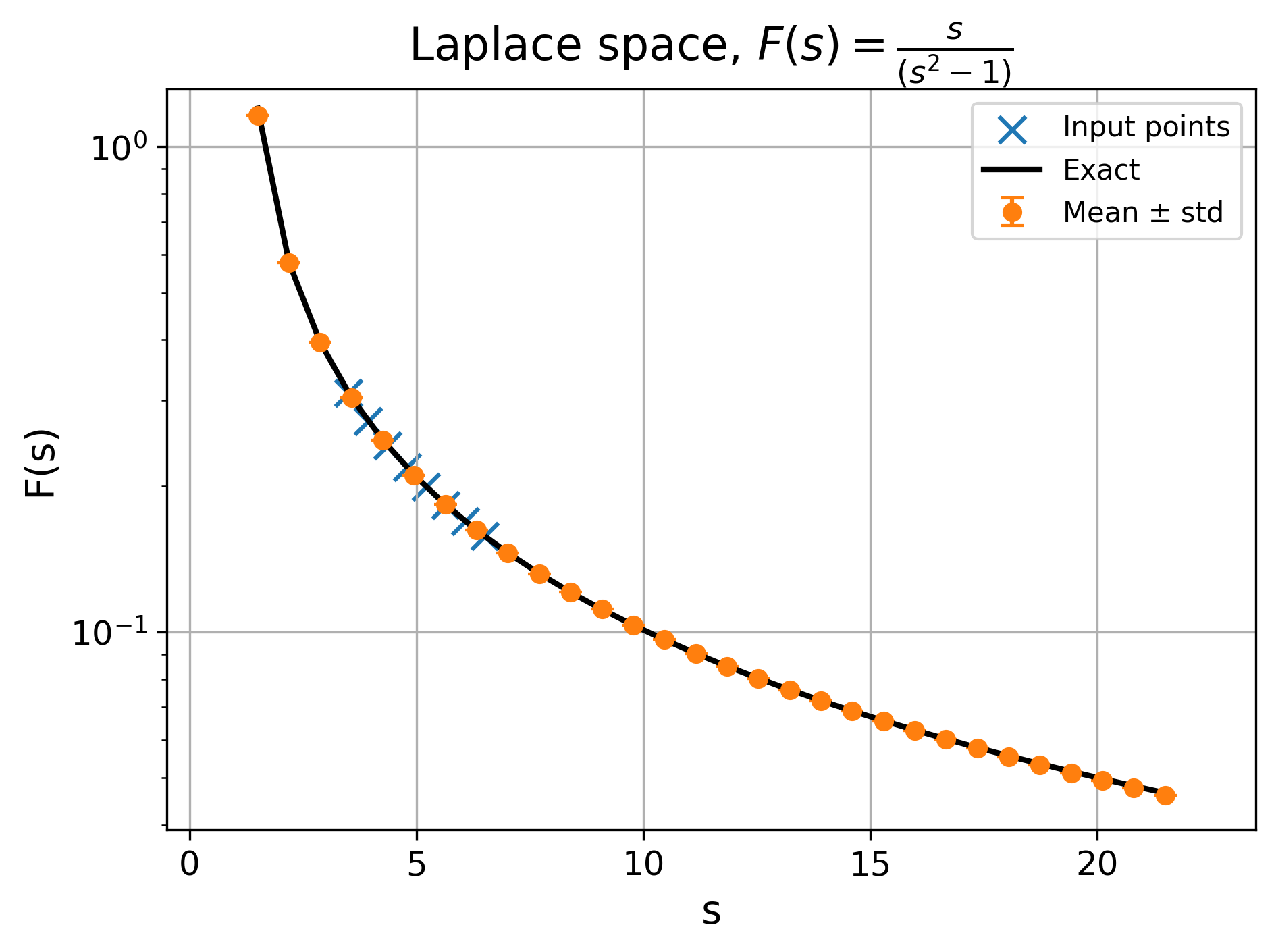}
    \end{subfigure}

    \caption{
    Numerical inversion and reconstruction tests for several analytic Laplace-space functions with known inverse transforms, performed in the absence of noise. For each column, the upper panel shows the reconstructed inverse function, while the lower panel shows the corresponding reconstruction of the Laplace-space function. The symbols and curves follow the same conventions as in Fig.~\ref{fig:laplace_test}.
    }
    \label{fig:laplace_test_additional}
\end{figure}

Increasing the number of input points improves resolution but also increases the condition number of the resulting linear system, highlighting the importance of scale selection and regularization. Similar behavior is observed for other analytic test functions, provided the reparameterization scale is chosen appropriately (see Fig.\ref{fig:laplace_test_additional}).

\subsection{Introducing Noise}

\begin{figure}[htbp]
    \centering
    %------------- Column 1 -------------
    \begin{subfigure}[t]{0.4\textwidth}
        \centering
        \includegraphics[width=\linewidth]{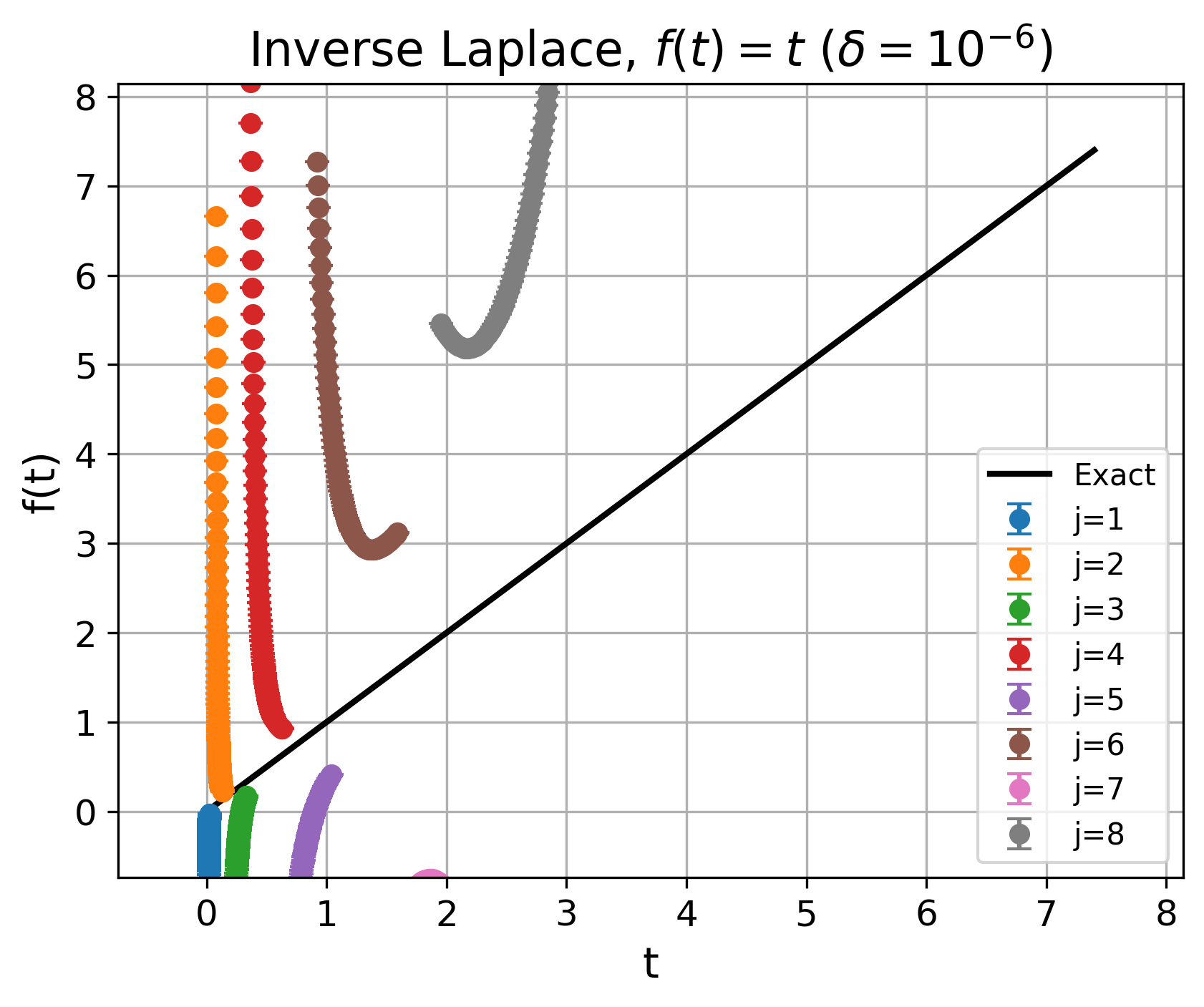}
        \includegraphics[width=\linewidth]{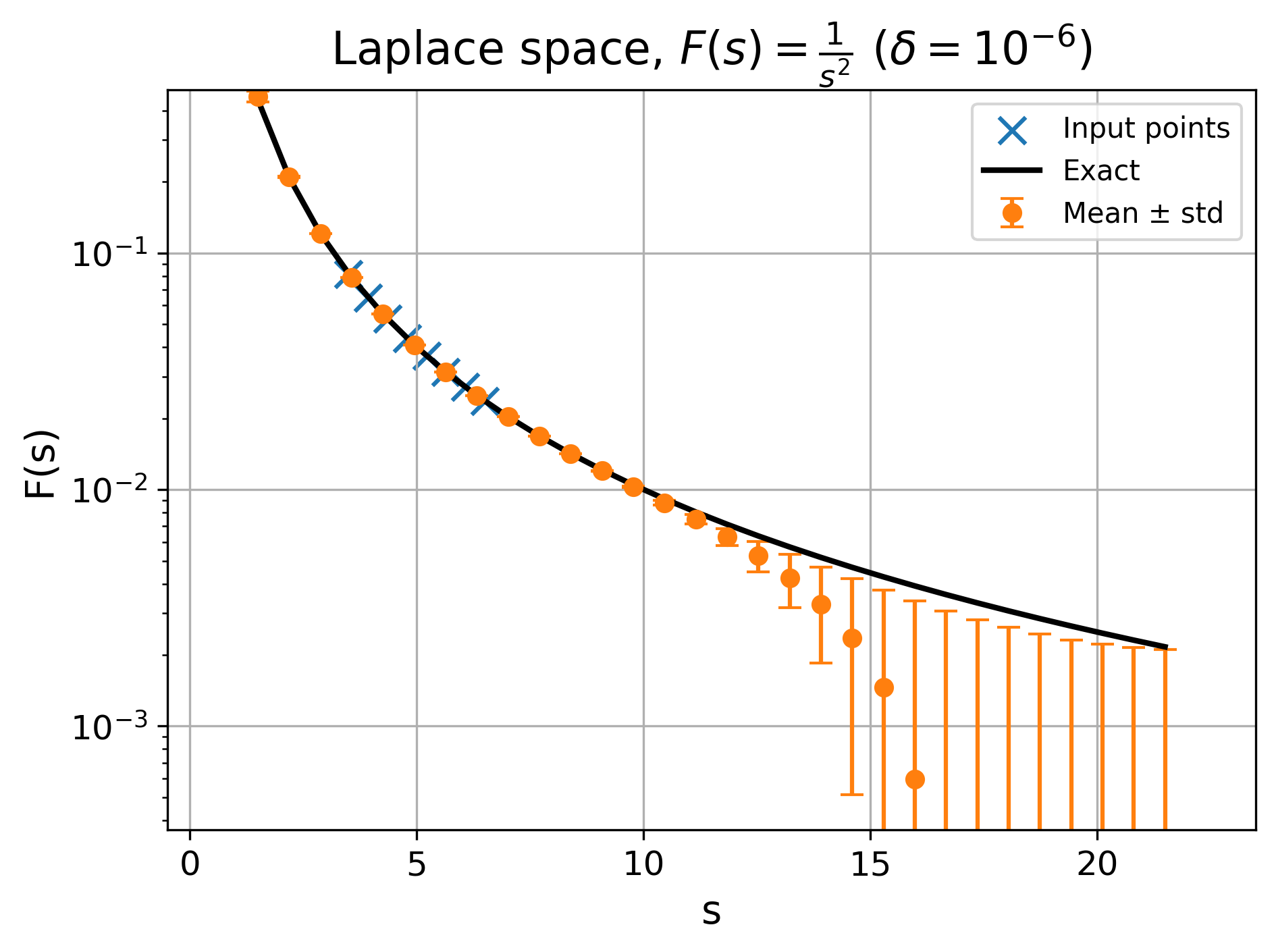}
    \end{subfigure}
    \hfill
    %------------- Column 2 -------------
    \begin{subfigure}[t]{0.4\textwidth}
        \centering
        \includegraphics[width=\linewidth]{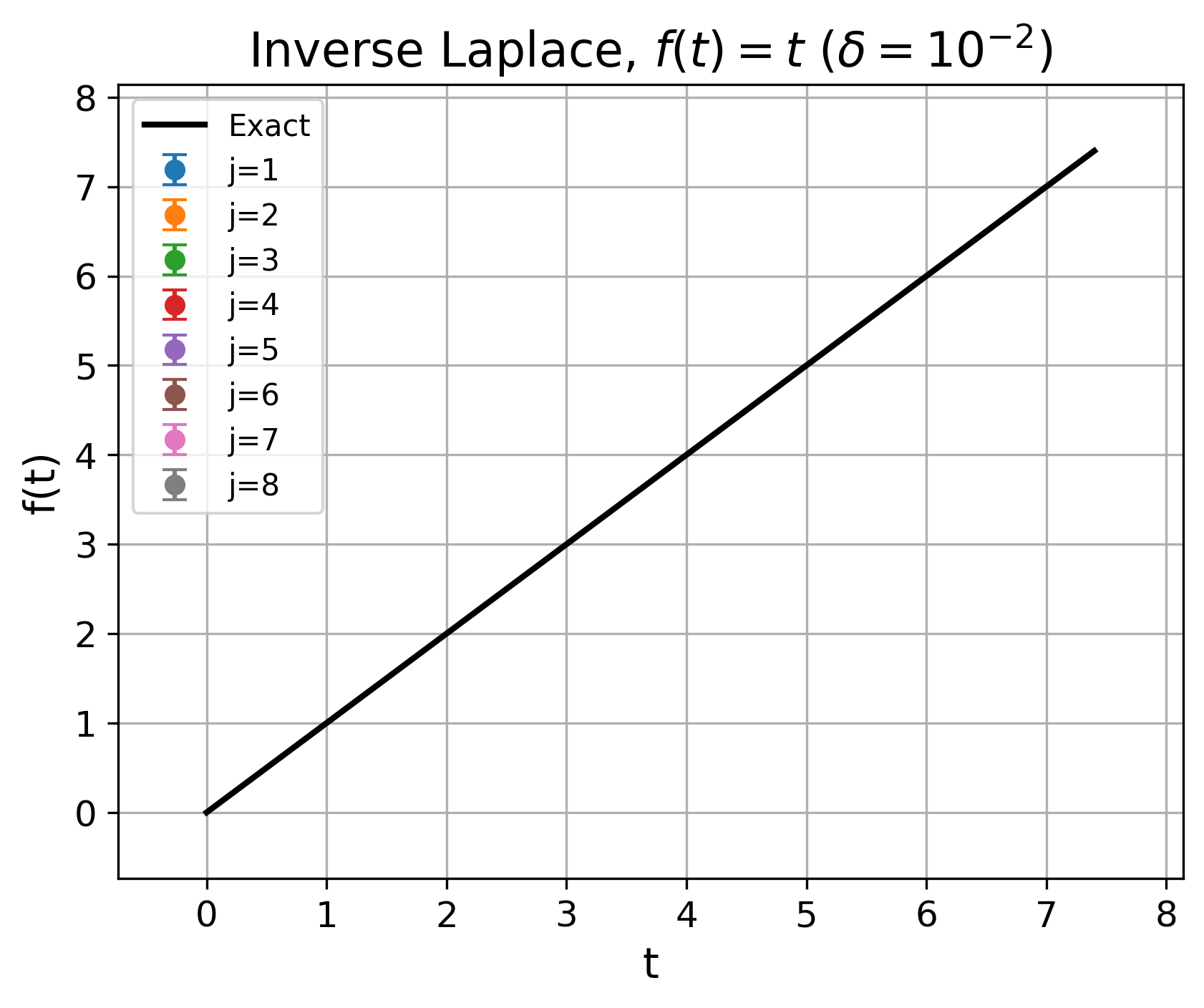}

        \includegraphics[width=\linewidth]{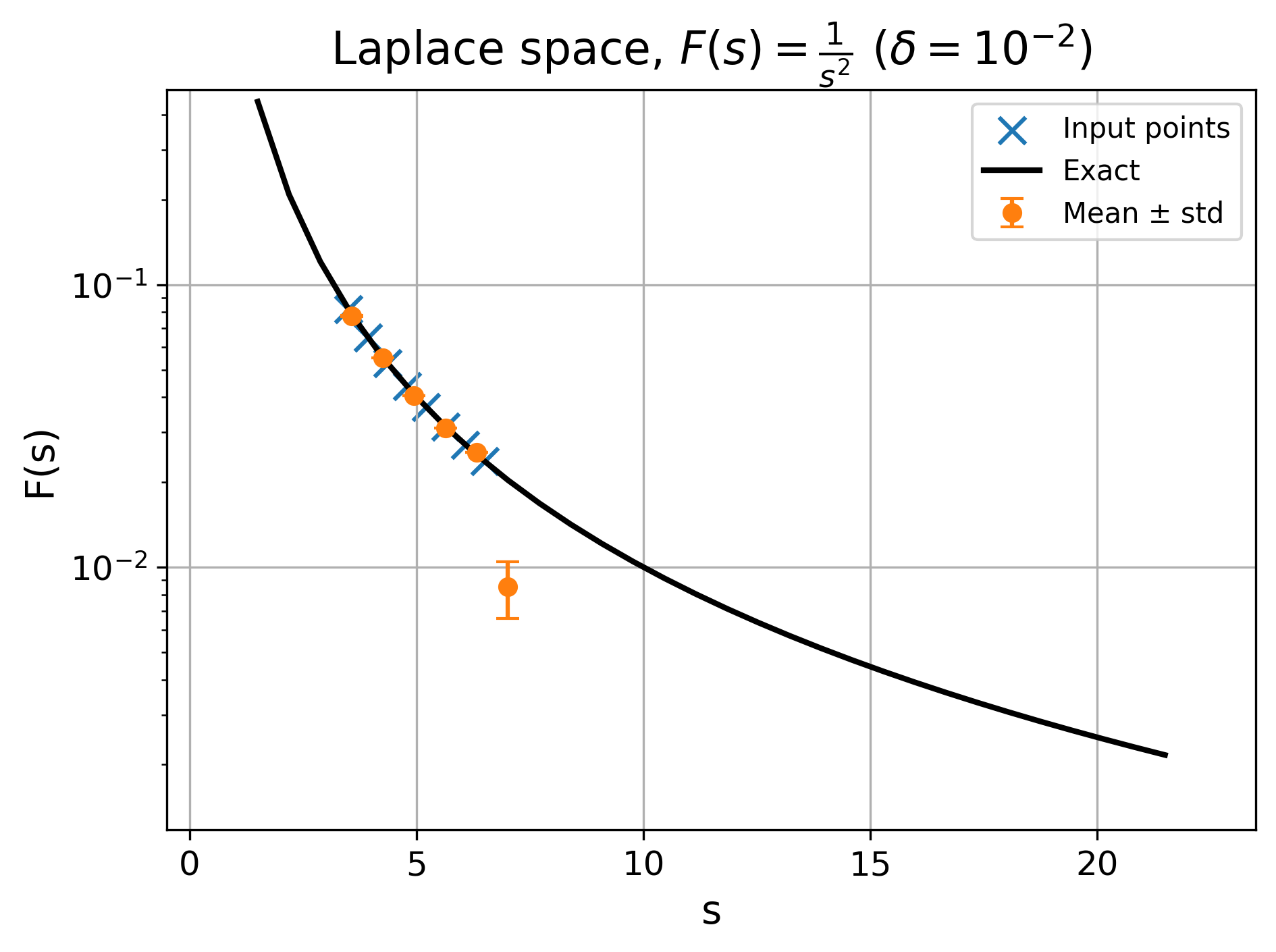}
    \end{subfigure}
    \caption{
    Numerical inversion and reconstruction for $F(s)=1/s^{2}$ with analytic inverse $f(t)=t$ and injected Gaussian noise, without additional denoising. The left column corresponds to relative noise level $\delta=10^{-6}$, while the right column shows results for $\delta=10^{-2}$ (Eq.~(\ref{eq:noise_injection})).
    }
    \label{fig:laplace_test_noise}
\end{figure}

To assess robustness, we inject Gaussian noise into the input data. Specifically, each data point $F(s_i)$ is added with noise drawn from a normal distribution with zero mean and standard deviation
\begin{equation}
\sigma_i = |F(s_i)|\,\delta,
\label{eq:noise_injection}
\end{equation}
where $\delta$ sets the relative noise level. As $\delta$ increases, reconstructions obtained at different scales diverge more strongly, indicating the need for additional stabilization. Fig.~\ref{fig:laplace_test_noise} illustrates how noise actually affects the previously discussed analytic example.

\subsubsection{Weighted Local Polynomial Smoothing}

\begin{figure}[htbp]
    \centering
    \includegraphics[width=0.65\textwidth]{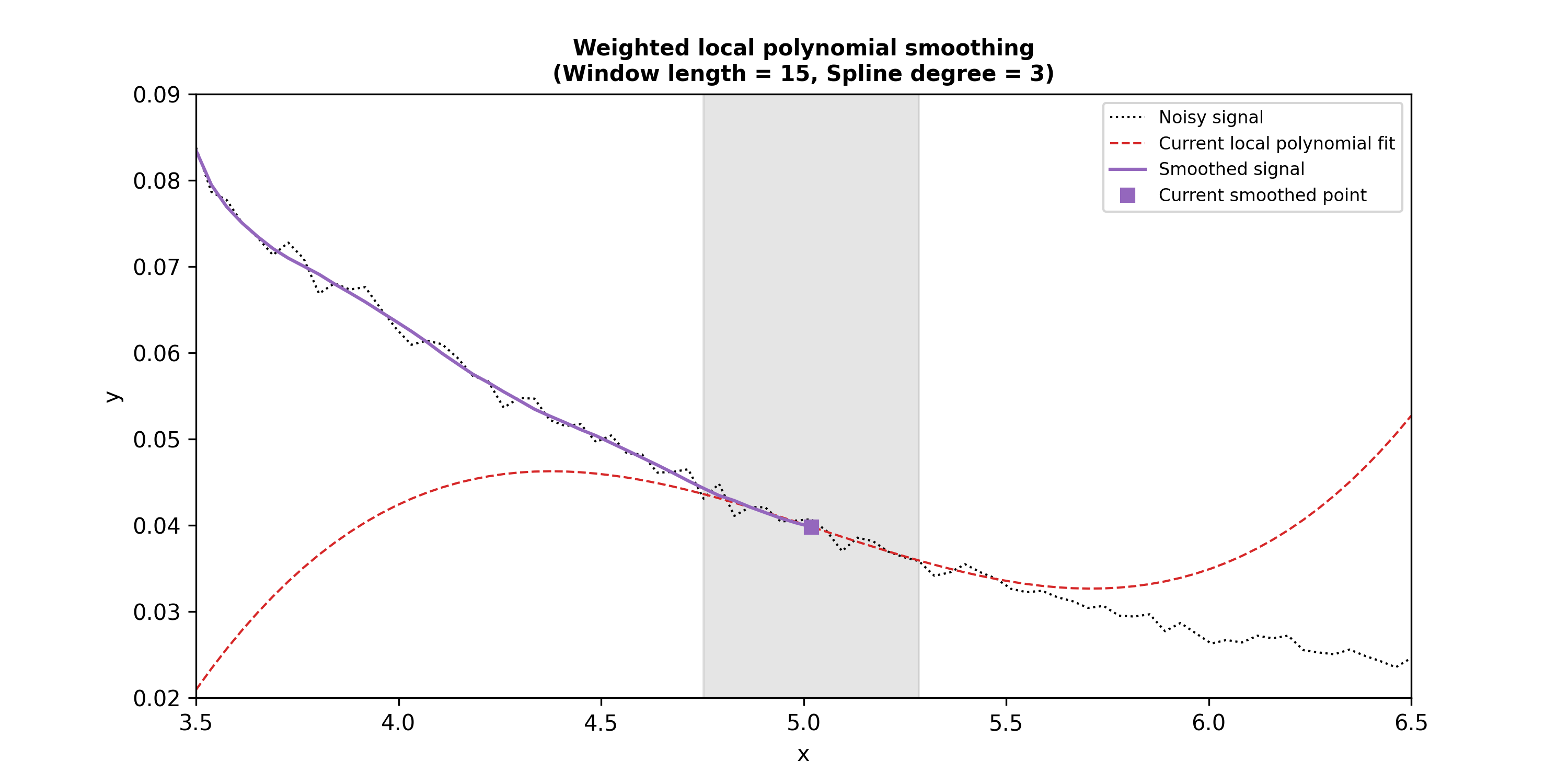}
    \caption{
    Illustration of weighted local polynomial smoothing. The black dotted curve shows the noisy signal, the red dashed curve the current local polynomial fit within the sliding window (shaded region), and the purple curve the resulting smoothed signal. The square marker denotes the currently smoothed point.
    }
    \label{fig:local_smoothing}
\end{figure}

As a first noise-mitigation step, we apply weighted local polynomial smoothing to the input data prior to inversion. For each point, a local neighborhood is selected and weighted with a Gaussian kernel centered on the point of interest. A low-order polynomial or spline is fitted to the weighted data and evaluated at the center. The window size controls the smoothing strength, while the polynomial order determines local flexibility. This preprocessing step significantly reduces noise amplification during inversion. This smoothing procedure can be better seen graphically in Fig.~\ref{fig:local_smoothing}.

\subsubsection{Denoising via Stochastic Optimization}

To further suppress noise-induced instabilities, we introduce a denoising strategy based on stochastic perturbations of the input data. Small random perturbations are applied, and the inversion is repeated across reparameterization scales. A fitness function is defined by quantifying the overlap between reconstructions obtained at neighboring scales, with the objective of minimizing discrepancies in overlapping regions and enforcing internal consistency. The optimization is carried out using the Covariance Matrix Adaptation Evolution Strategy (CMA-ES) \cite{HansenCMAES}. This procedure effectively suppresses noise-driven fluctuations while preserving the underlying signal.

\begin{figure}[htbp]
    \centering

    \begin{subfigure}[t]{0.48\textwidth}
        \centering
        \includegraphics[width=\linewidth]{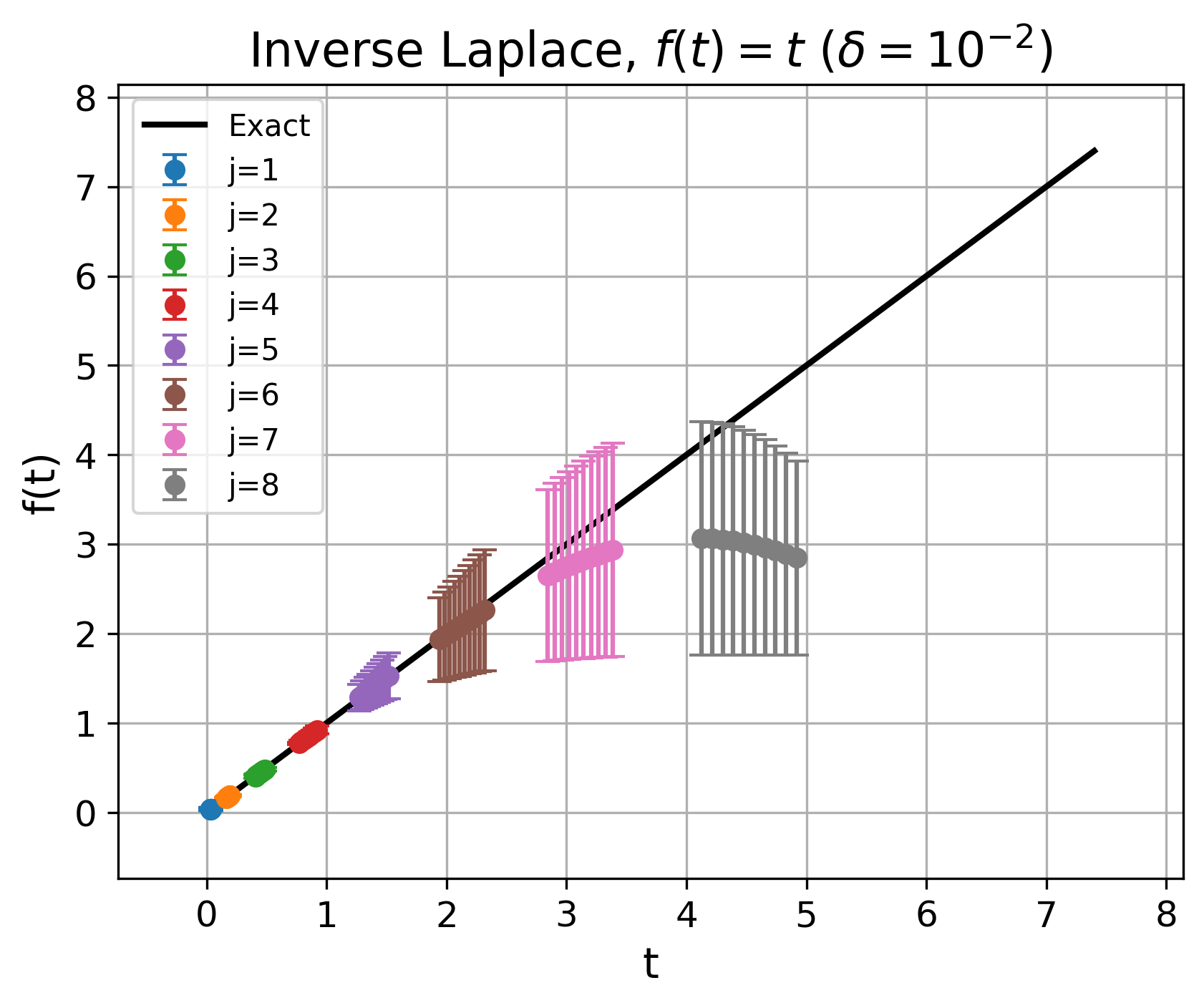}
        \caption{
        Inverse Laplace reconstruction of the analytic function $f(t)=t$, after applying smoothing and stochastic denoising. Colored markers indicate reconstructed values at the Gauss--Laguerre quadrature nodes $t_0 t_j$ obtained from 10 independent optimization runs, with error bars showing the spread. The black curve represents the exact solution.
        }
        \label{fig:inverse_laplace_denoise}
    \end{subfigure}
    \hfill
    \begin{subfigure}[t]{0.48\textwidth}
        \centering
        \includegraphics[width=\linewidth]{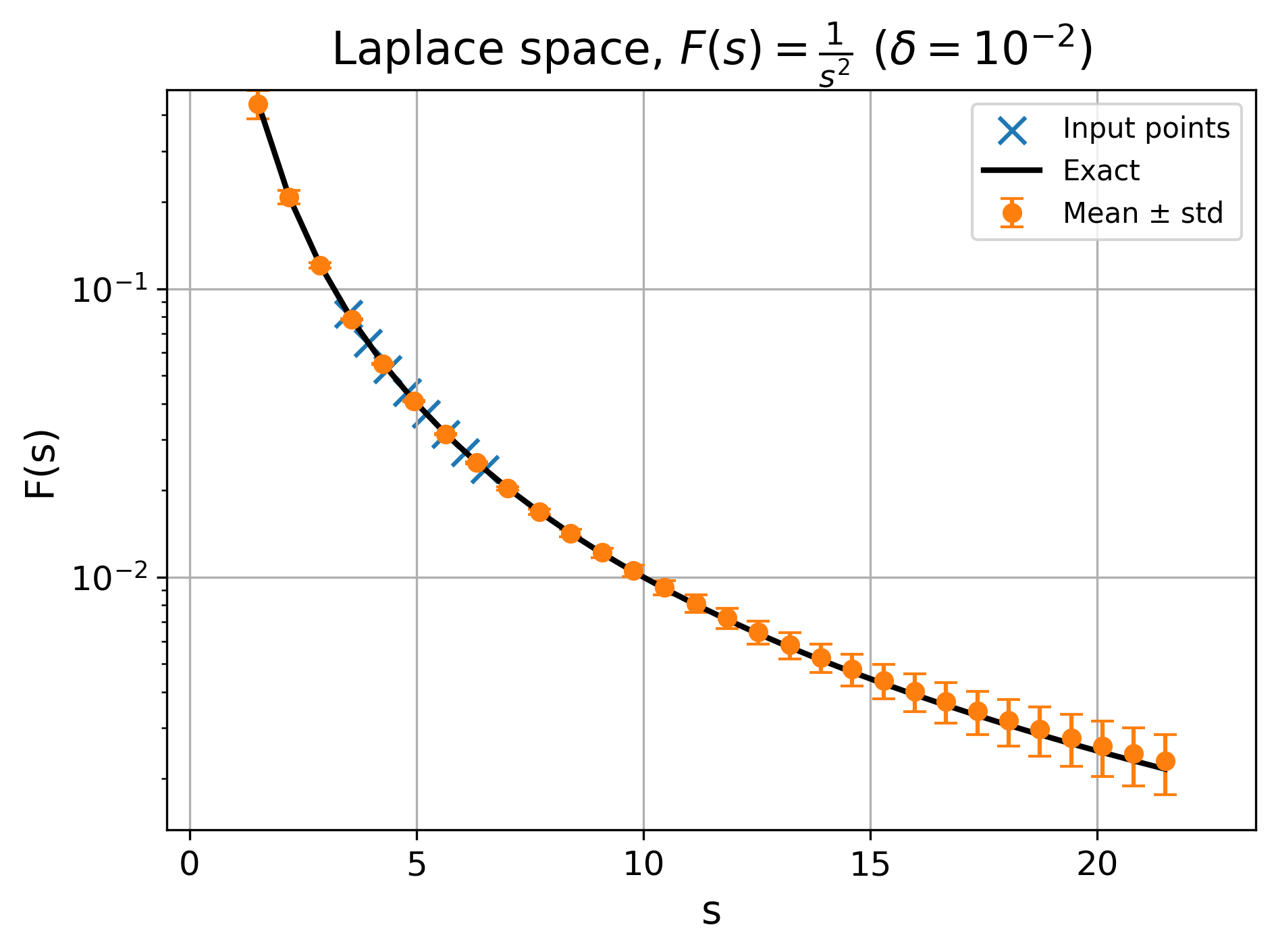}
        \caption{
        Reconstruction of the Laplace-space function $F(s)$ from the denoised inverse solution shown in panel~\ref{fig:inverse_laplace_denoise}. Blue crosses indicate the original noisy input data points, orange circles show reconstructed and extrapolated averaged values with error bars, and the black curve shows the exact result.
        }
        \label{fig:reconstruction_extrapolation_denoise}
    \end{subfigure}

    \caption{
    Numerical inversion and reconstruction test for $F(s)=1/s^{2}$ with analytic inverse $f(t)=t$, with injected noise and denoising applied.
    }
    \label{fig:laplace_test_denoise}
\end{figure}

Figures~\ref{fig:inverse_laplace_denoise} and \ref{fig:reconstruction_extrapolation_denoise} show the resulting inverse and reconstructed Laplace-space function for the noisy test case. Compared to the unsmoothed inversion in Fig.\ref{fig:laplace_test_noise}, the combined smoothing and stochastic denoising yield stable reconstructions and reliable extrapolation.

\section{Application to Spectral Density Reconstruction}

\subsection{Lattice Spectral Representation}

In the infinite-time limit, a lattice correlator is related to the spectral density by
\begin{equation}
C_L(t) = \lim_{T \to \infty} C_{LT}(t) = \int_0^{\infty} dE \, e^{-tE} \rho_L(E).
\end{equation}
In practice, finite lattice spacing, limited volume, and statistical noise make the inversion from $C_L(t)$ to $\rho_L(E)$ ill-posed. At finite volume, the spectral density consists of a discrete set of energy levels with positive weights. The smearing introduced by the quadrature-based inversion should therefore be interpreted as defining an effective, resolution-limited spectral density rather than a direct estimator of the finite-volume spectrum.

\subsection{Gauss--Legendre Quadrature}

Although the spectral density formally extends to infinite energy, high-energy contributions are exponentially suppressed in the Laplace transform.  For numerical purposes, we therefore employ Gauss--Legendre quadrature over a finite interval $[a,b]$ that captures the dominant contributions:
\begin{equation}
C_L(t_i) \approx \int_a^{b} dE \, e^{-t_iE} \rho_L(E) \approx \sum_{j=1}^{n} \tilde{w}_j e^{-E_j t_i} \rho_L(E_j),
\end{equation}
where $E_j$ and $\tilde{w}_j$ are the Legendre nodes and weights. This approximation can then be inverted using the quadrature-based, multi-scale method introduced previously. Using Gauss--Legendre quadrature ensures that quadrature nodes are concentrated where $\rho_L(E)$ is non-negligible. In contrast, Gauss--Laguerre quadrature may place many nodes in regions with negligible spectral weight, making it less efficient for lattice spectral densities.

The bounds $[a,b]$ are chosen such that contributions outside the interval are negligible over the Euclidean-time range considered; modest variations of $a$ and $b$ are found not to affect the reconstructed correlators.

\subsection{Mock Data Tests}

To validate our inversion procedure, we generate mock spectral densities consisting of ten discrete energy levels, evenly spaced in the interval $[2 m_\pi, 8 m_\pi]$ ($m_\pi = 140$~MeV). This construction follows the mock-data setup of Ref.~\cite{DelDebbio2025}. The corresponding correlators are then computed at 64 Euclidean time slices:
\begin{equation}
C(t) = \sum_{n=0}^{9} w_n \, e^{-E_n t}.
\end{equation}
The weights $w_n$ are sampled from a correlated multivariate Gaussian distribution:
\begin{equation}
K_{nn'} = \kappa \exp\Big[-\frac{(E_n-E_{n'})^2}{2\epsilon^2}\Big],
\end{equation}
with correlation length $\epsilon=0.01\,\Delta E$ and variability $\kappa = 0.1\, m_\pi$. For each of these sets of weights, we inject statistical noise into the corresponding correlator by using a covariance matrix computed in lattice QCD, for a vector-vector two-point correlation function of light-quark mesons.

\begin{figure}[htbp]
    \centering

    \begin{subfigure}[t]{0.48\textwidth}
        \centering
        \includegraphics[width=\linewidth]{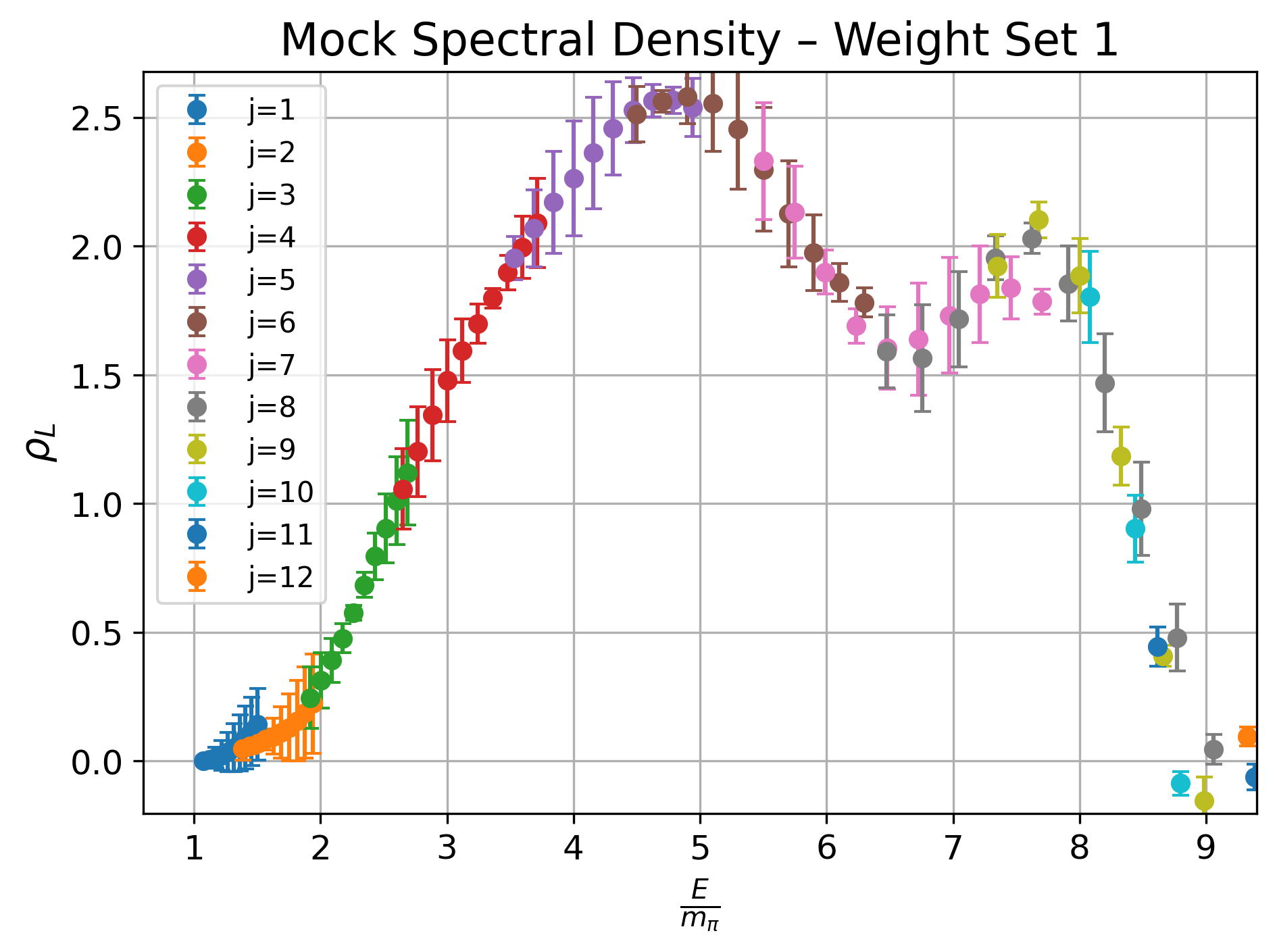}
        \caption{
        Reconstruction of a smeared spectral density from mock lattice correlator data using the quadrature-based inversion and denoising procedure. Results are shown for 10 independent optimization runs, illustrating the stability of the reconstructed spectral features.
        }
        \label{fig:inverse_laplace_mock_data}
    \end{subfigure}
    \hfill
    \begin{subfigure}[t]{0.48\textwidth}
        \centering
        \includegraphics[width=\linewidth]{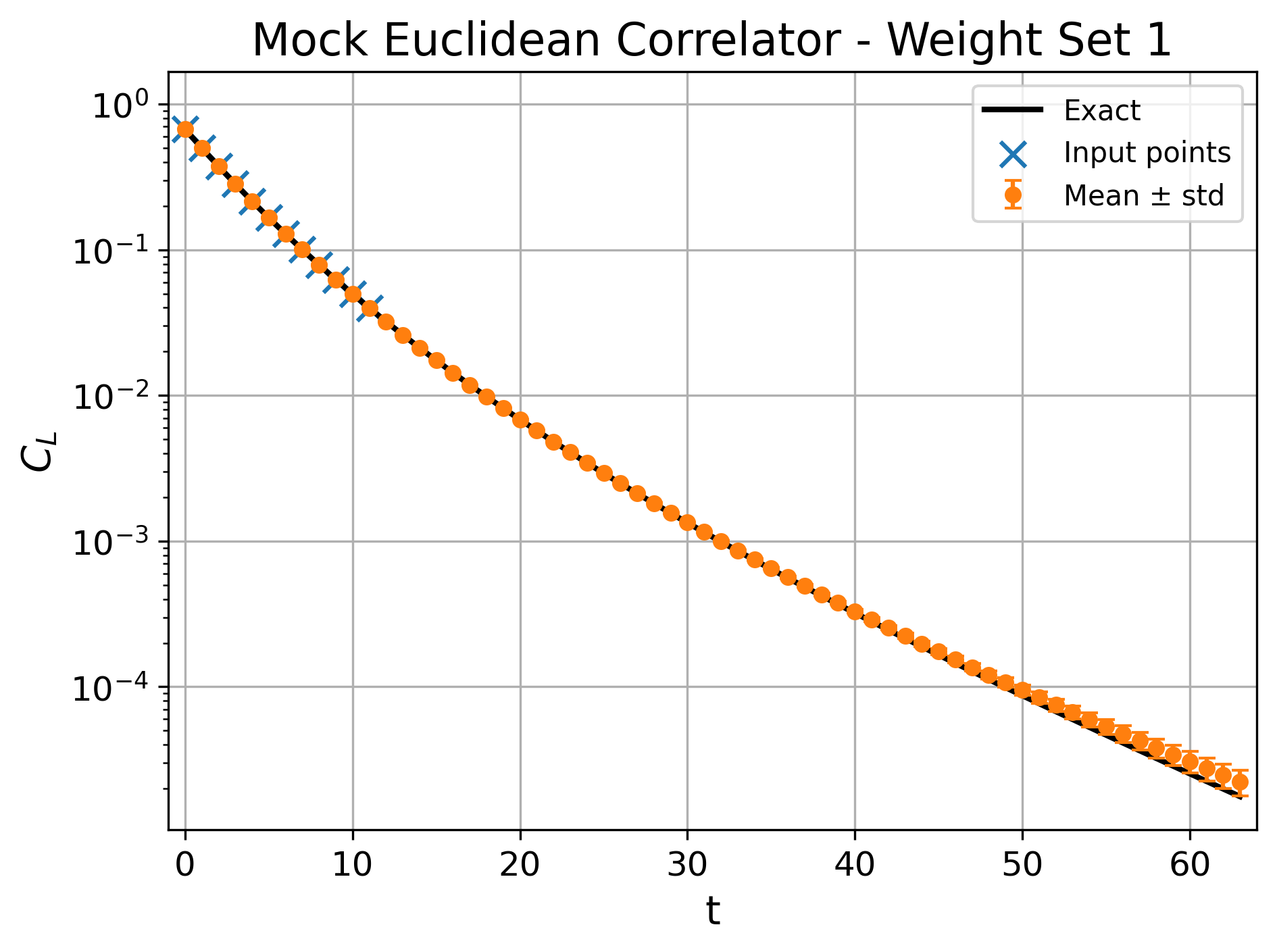}
        \caption{
        Euclidean correlator corresponding to the reconstructed spectral density in panel~\ref{fig:inverse_laplace_mock_data}. The black curve shows the original correlator (mostly obscured), blue crosses indicate the first 12 time slices used as input for the inversion, and orange circles represent the reconstructed correlator values.
        }
        \label{fig:reconstruction_extrapolation_mock_data}
    \end{subfigure}

    \caption{
    Numerical inversion and reconstruction for Weight Set~1 of the mock data using the quadrature-based denoising procedure.
    }
    \label{fig:laplace_mock_data}
\end{figure}

\begin{figure}[htbp]
    \centering

    %------------- Column 1 -------------
    \begin{subfigure}[t]{0.4\textwidth}
        \centering
        \includegraphics[width=\linewidth]{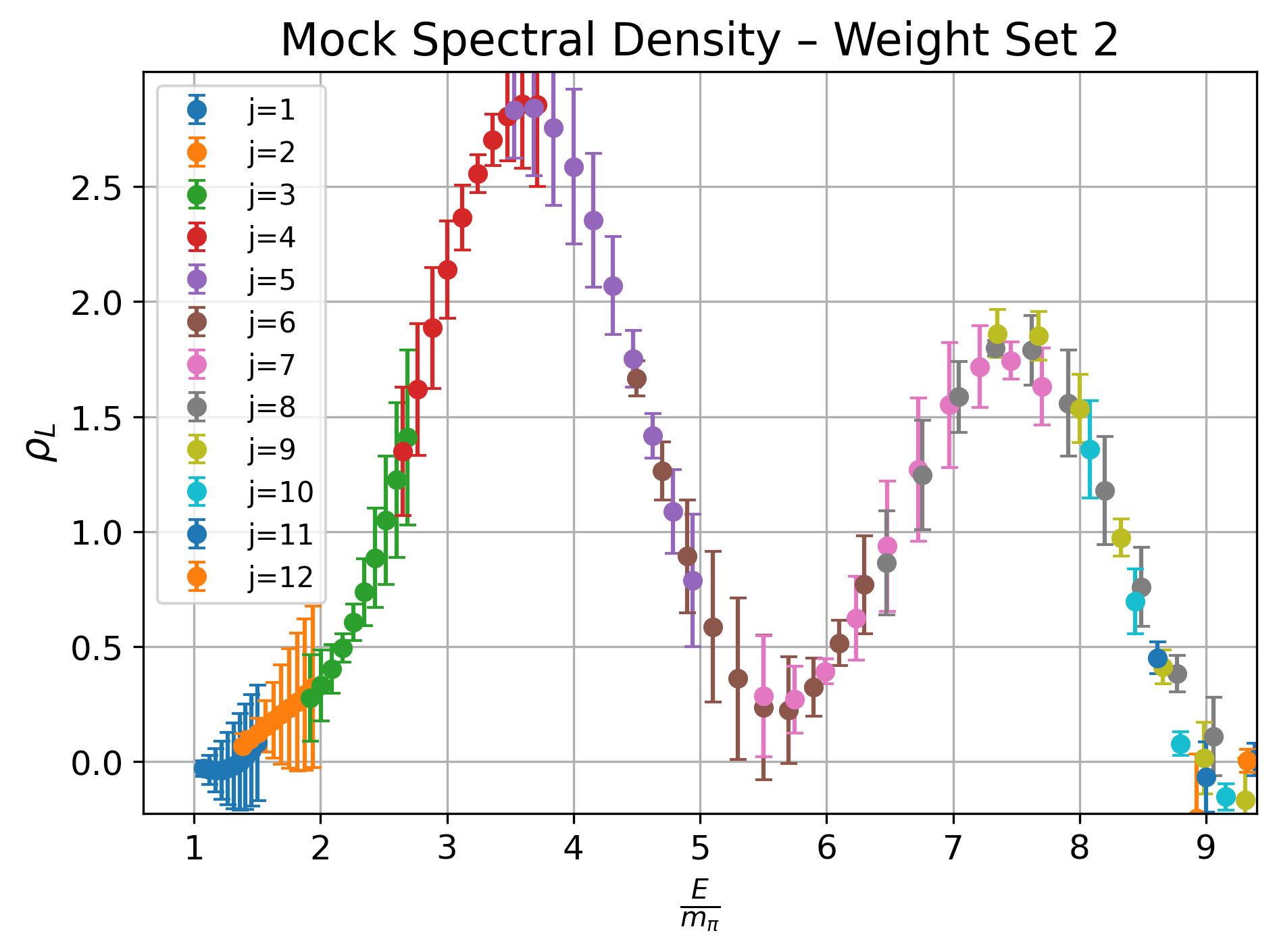}
        \includegraphics[width=\linewidth]{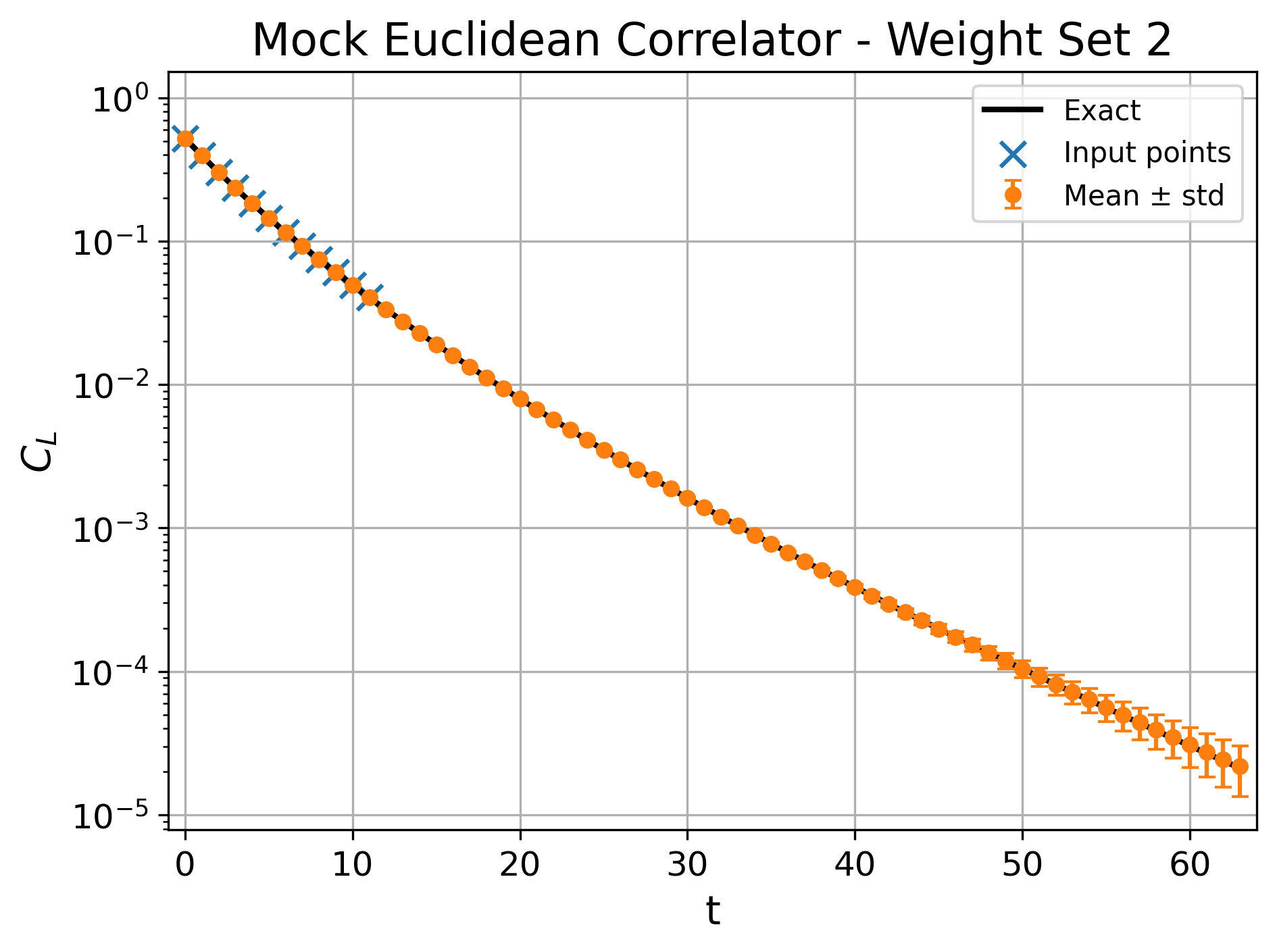}
    \end{subfigure}
    \hfill
    %------------- Column 2 -------------
    \begin{subfigure}[t]{0.4\textwidth}
        \centering
        \includegraphics[width=\linewidth]{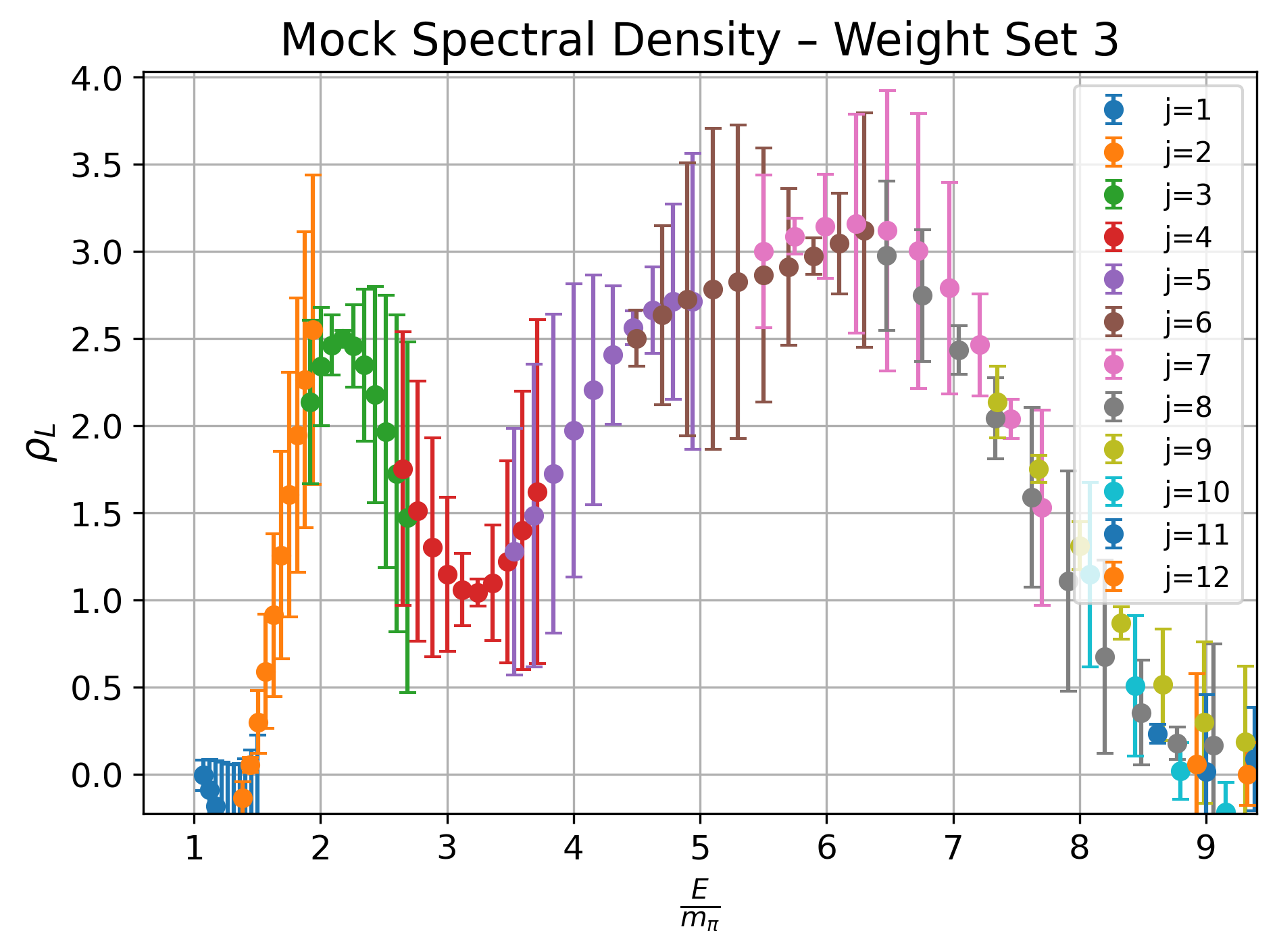}
        \includegraphics[width=\linewidth]{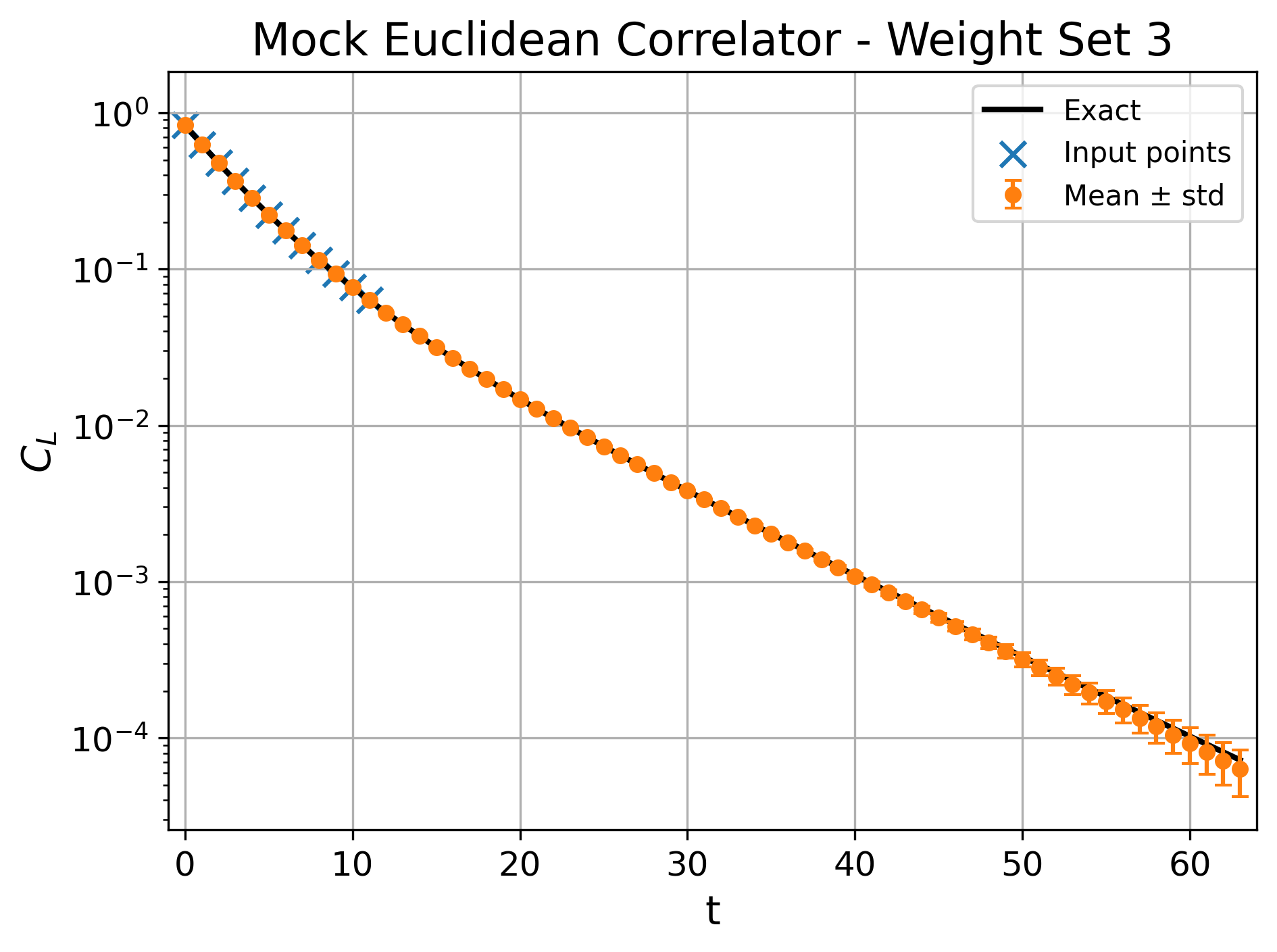}
    \end{subfigure}
    \caption{
    Numerical inversion and reconstruction for additional weight sets of the mock data. For each column, the upper panel shows the reconstructed smeared spectral density, and the lower panel shows the corresponding correlator reconstruction. The symbols and curves follow the same conventions as in Fig.~\ref{fig:laplace_mock_data}.
    }
    \label{fig:laplace_mock_data_additional}
\end{figure}

Applying Gauss--Legendre quadrature together with the multi-scale denoising method, we obtain stable reconstructions of the smeared spectral density. Only the first 12 Euclidean time slices are used as input in order to mimic realistic lattice analyses, where late-time data are typically dominated by noise and carry limited independent information. The agreement between the reconstructed and original correlators at large Euclidean times provides a non-trivial validation of the inversion procedure, since these time slices were not included in the input data (see Figs.~\ref{fig:inverse_laplace_mock_data} and \ref{fig:reconstruction_extrapolation_mock_data} for Weight Set~1). Additional reconstructions for mock data with varied spectral weights are shown in Fig.~\ref{fig:laplace_mock_data_additional} (Weight Sets~2 and~3), further demonstrating the robustness of the inversion procedure.

\section{Conclusions and Outlook}

In this study, we have developed a quadrature-based framework for the numerical inversion of Laplace transforms from discrete and noisy data. By combining Gauss-type quadratures with a multi-scale reparameterization strategy, the method closely follows the structure of the underlying integral transform and yields stable, accurate inversions. Tests on toy models highlight the robustness of the approach in the presence of noise and limited data. Moreover, tests on mock lattice correlators demonstrate encouraging reconstructions of spectral densities, paving the way for applications to actual lattice QCD correlators.

Future work will focus on several directions. First, the method will be tested on additional synthetic benchmark systems to further assess its accuracy and limitations. Second, applications to actual lattice correlators, such as vector and pseudoscalar channels from existing ensembles, will be pursued. Further improvements include developing more robust criteria for selecting the reference scale $t_0$ in noisy environments using additional statistical metrics, incorporating more correlator data through suitable regularization strategies, and gaining a deeper understanding of the resulting smeared spectral densities and the impact of finite temporal ranges on the inversion. Finally, systematic comparisons with established reconstruction methods will be carried out to quantitatively assess performance and reliability.

\section*{Acknowledgements}

This work was supported by INFN under the research project (iniziativa specifica) \textit{QCDLAT}. The authors thank Alessandro De~Santis for providing the covariance matrix used to generate the mock datasets employed in this study.

\end{document}